\documentclass[a4paper,12pt]{article}

\usepackage{amsmath,amssymb}

\usepackage{amsmath}
\usepackage[mathscr]{eucal}

\usepackage{lmodern}
\usepackage{extarrows}
\usepackage{rotating}
\usepackage{caption}
\usepackage{float}

\usepackage{graphicx}
\usepackage{xfrac}

\usepackage{xcolor}

\begin{document}

\begin{center}

\section*{Quantum Vortices in a Boundary Layer: \\ New  Results and Perspectives }

\vspace{3mm}

{S.V. Talalov}

\vspace{3mm}

{\small Institute of Digital Technology, Togliatti State University, \\ 14 Belorusskaya str.,
 Tolyatti, Samara region, 445020 Russia.\\
svt\_19@mail.ru}

\begin{abstract}
Here, we investigate the motion of a thin circular quantum vortex filament near the infinite planar surface.
  The fluid surrounding this surface moves with a non-zero velocity $\bf{v}$, which is parallel to the surface.  We study the specific features of this quantum  system and show that they are quite suitable for the boundary layer theory.
	The developed model  allows us to calculate the vortex energy spectrum, $E = E({\bf p})$, where ${\bf p}$ is the total  momentum of a vortex ring. 				
		We have demonstrated that this function  has complex non-trivial dependence    on the velocity $\bf{v}$. 
		It is stated that the inverse effective mass of a vortex under consideration is of a tensorial nature. In certain quantum states, the system shows the possibility of both negative and positive effective mass existing.  
	This study employs a novel quantization method for classical closed vortex filaments, developed by the author earlier.
\end {abstract}


{\bf keywords:} ~   quantum vortices,  ~negative effective masses, boundary layer.
 
\end{center}


 {\bf PACS numbers:}   47.10.Df    ~~47.32.C

\vspace{5mm}

\subsection*{1. Introduction}

~~~The boundary layer theory is a well-established concept in classical fluid dynamics. The early stages of this theory's development, as well as its different results, can be seen, for example, in the book \cite {Schli} (see also the book \cite{Saffm} where the behavior of the classical vortices near the surfaces was explored). 
The mathematical aspects of boundary layer theory were studied in detail in the book \cite {Lomov}.
Since these books were edited, many issues in this field were investigated concerning classical fluid flows. We are not aiming to review these results here, because we will be considering the quantum case instead. 
For some time, it has been believed that there is no boundary layer in superfluids.
But recent research \cite{StPaBa} has shown that this statement needs to be revised. 

In this study, we construct a  mathematical model that describes the evolution of a quantum circular vortex filament near an infinite planar surface. 
We also explore how the findings apply to the process of boundary layer formation.
The suggested model is based on the new approach developed earlier by the author for vortex quantization \cite{Tal24_1,Tal25_J}. In these and other author's  papers, the background and motivation for this non-traditional approach were discussed at length.
Instead of repeating details, we highlight the essential and innovative aspects of our proposed method.

Firstly, we introduce quantum vortices into a theory in a special non-standard way.
In contrast to the prevailing view, our approach proposes that vortices are quantized classical  dynamical systems, rather than topological defects.
 Among other things, our findings differ from the standard in that the range of quantized circulations, $\Gamma$, is wider than the well-known equally spaced levels described by the equation $\Gamma=\hbar n/\mu_H$. (where $n = 1, 2, \ldots$).
 The previous author's papers thoroughly explored the reasoning behind this significant finding.

Secondly, we use a group-theoretical framework to determine the total energy of a   vortex filament with very small but non-zero core radius ${\sf a}$. 
 To implement this program, we introduced a set of independent Hamiltonian variables, enabling a vortex loop to be represented as a particle with internal degrees of the freedom. 
Here, we introduce the centrally-extended Galilean group $\widetilde{\mathcal{G}}_3$. This group serves as the space-time symmetry group  of our theory. 
Lee algebra of the  group $\widetilde{\mathcal G}_3$ has three Cazimir functions:  
		 \[ {\hat C}_1 = \mu_0 {\hat I}\,,\quad 
  {\hat C}_2 = \Bigl({\hat M}_i  - \sum_{k,j=x,y,z}\epsilon_{ijk}{\hat P}_j {\hat B}_k\Bigr)^{\!2} 
  \quad {\hat C}_3 = \hat H -  \frac{1}{2\mu_0}\sum_{i=x,y,z}{\hat P}_i^{\,2}\,,\]                        
       where        ${\hat I}$ is the unit operator,     ${\hat M}_i$,   $\hat H$,  ${\hat P}_i$         and  ${\hat B}_i$  ($i = x,y,z$)
        are the respective generators of rotations, time and space translations and Galilean boosts, value $\mu_0$ is a central charge. 
 As usual, the function ${\hat C}_3$ is understood as the ''internal energy of the particle''. 
The central charge $\mu_0$ can be seen as a sort of conditional ''bare mass''. 
This method allows us to bypass intricate and sometimes ambiguous hydrodynamic calculations involving vortex filaments with very thin cores.
Let us recall here  Donnelly's insightful comment: {\it''\dots  considering how small the vortex core in helium II is, i.e., of order an angstrom, it would seem that one either ought to know how it is constructed or one ought to find a way to ignore it. Unfortunately neither goal has been achieved''} \cite{Donn}.	
Thus, we propose the following formula as the fundamental equation for vortex energy:
\begin{equation}
		\label{E_general}
 {E}_{cl} ~=~ \frac{{\bf p}^{\,2}}{2\mu_0} ~+~ {\hat C}_3(\varpi, \chi, \dots)\,.
\end{equation}
Here, the vector ${\bf p}$ represents vortex momentum, while the variables $\varpi, \chi, \ldots$ are a set of "internal" variables. 
In section 3, we will elucidate the concept of vortex energy in detail.  As a consequence,  we will introduce the concept of the  ''effective mass'' for the circular vortex filament.

Thirdly, selecting independent Hamiltonian variables enables us to apply the many-body theory to describe interactions among vortices. The author has addressed this subject before, but it doesn't align with the current paper's primary focus.
As regards the planned outcomes of our study, they are:
\begin{itemize}
\item We intend to explain the concentration of vortices near the boundary plane;
\item We will calculate the vortex energy as a function of the vortex momentum. We will also calculate and analyze the inverse effective mass tensor for the circular vortex filament
\item Although our model does not explore the reasons for the vortex appearance\footnote{The paper  \cite{StPaBa} attributes this phenomenon to surface irregularities.} 
  near the surface, we will formulate criteria to determine when vortices will occur;
	\item We will explore a simplest multi-vortex system near a plane, concentrating on how the boundary layer develops.
\end{itemize}

\subsection*{2. The physical system under study}

Let's define the physical system we'll explore here.
Generally speaking, we consider the circular-shaped, thin  quantum  vortex filaments that evolve near an infinite plane. As a classical object before quantization, filament's behaviour is governed by well-known differential equation of Local Induction Approximation (LIA), augmented by term capturing flow within vortex core. Before  writing it, we should  make the important notes and  outline our desired outcomes.

\begin{enumerate}
\item Because the area of applicability of LIA is quite limited, we should consider the possibility of its generalization. This generalization has been made in the paper \cite{Tal24_1}.  But to make our construction more visual and simpler, we consider only the LIA equation with the mentioned additional term;
\item Of course, we can extend our theory by considering small perturbations to the ring\footnote{The author explored this topic in their early writings.}. But in our approach, the quantum filaments shaped like circles can be seen as building blocks, or "atoms," for creating closed quantum filaments of any size and shape. Corresponding theory has been developed in the paper \cite{Tal26_1};
\item The quantum states that will be constructed, admit the states with some flow in the vortex core as well as the states without it. The author recognizes diverse viewpoints on this matter. Maybe, the point of view given above in the quote by Donnelly justifies including such a flow in our consideration;
\end{enumerate}

To build our model,   we use the following constants:
the surrounding  fluid density $\rho_0$ and the speed of sound in this fluid $v_0$. 
Besides this,  we also need to introduce the length constant $R_f$.  This constant can be related to the size of a stable molecular cluster, intermolecular distance and so on. Here, we don't specify it, but we assume that the constant $R_f >> {\sf a}  > 0$ defines a certain minimal radius for the vortex ring.  Unlike the constant $R_f$, the core radius $\sf a$  remains undetermined here.

 Further, to simplify the formulas, we will use the auxiliary constants such as 
$  t_0 = R_f/v_0$  and  ${\mathcal E}_0 = \mu_0 v_0^2$.
   The  constants $t_0$ and  ${\mathcal E}_0$ define the  time and energy scales in constructed classical  theory.

 It is necessary  remind   that the standard LIA equation is deduced \cite{Saffm} for the ''evolution parameter'' $t^*=t\Gamma/4\pi$, where $t$  means  the actual (physical) time and $\Gamma$ means the circulation.
Exploring the dynamics of the vortex ring, it will be more convenient  to use dimensionless parameters $\tau$ and $\xi$ instead of the ''real time'' $t$ and the natural parameter (arc length of the curve) ${\sf s}$:
\begin{equation}
        \label{tau_s}
 \tau    ~=~  t^* /  R^{\,2}\,,	\qquad \quad \xi ~=~ {\sf s}/{R}\,, \qquad
\xi \in [0,2\pi]\,,
 \end{equation}
where value $R$ means the ring radius. 
If we introduce the projective vector  ${\mathfrak r}\nolinebreak = \nolinebreak {\bf{r}}/R$ instead of the radius-vector ${\bf{r}}$, the convenient  form of the LIA equation will be as follows:

\begin{eqnarray}
        \label{LIE_pert}
        \partial_\tau {\mathfrak r}(\tau ,\xi)  & = &
        \alpha\, \Bigl(\partial_\xi{\mathfrak r}(\tau ,\xi)\times\partial_\xi^{\,2}{\mathfrak r}(\tau ,\xi)\Bigr)   ~+ \nonumber \\
				~~ & + & \omega\,\Bigl(2\,\partial_\xi^{\,3}{\mathfrak r}(\tau ,\xi) ~+~ 
        {3}\,\bigl\vert\, \partial_\xi^{\,2}{\mathfrak r}(\tau ,\xi)\bigr\vert^{\,2}\partial_\xi{\mathfrak r}(\tau ,\xi)\Bigr)\,,
				        \end{eqnarray}
where the values $\alpha$   and  $\omega$ \label{omega}    are finite dimensionless constants.
In total, the parameter $\alpha \propto \ln{\sf a} + const$
 arises as a consequence of a certain regularization process when the LIA equation is derived.  As regards the
 parameter $\omega$, it  is determined by the velocity of the fluid flow  $\Phi_\omega$ in the vortex core.  The book \cite{AlKuOk} thoroughly derives this equation and explicitly expresses the parameter $\omega$ using conventional physical values which are connected with core flow.  
{~ We will assume that both $\alpha$ and $\omega$ take non-zero values in this study.}

We use the following exact solution to the  Eq.(\ref{LIE_pert}):
\begin{equation}
        \label{our_sol}
 {\mathfrak r}(\tau ,\xi) ~=~ \Bigl(\, {q_x}/{R} ~+~  \cos(\xi +\phi)\,,\quad {q_y}/{R} ~+~  \sin(\xi +\phi  )\,, \quad {q_z}/{R} ~+~ \alpha \tau \,\Bigr)\,, 
\end{equation}
where  the  variable  $\phi \equiv \phi(\tau) =  \phi_0 +  \omega\tau$ 
 defines the rotation of the circular curve (\ref {our_sol}).  This rotation  models the flow into the core of the vortex.
As regards the vector $q = (q_x,q_y,q_z)$, it determines the center of the circle. 

Here and further, we will use the indices $(x, y, z)$ for the Cartesian coordinate system, which is naturally associated with solution (\ref{our_sol}). In this system, the $x0y$ plane coincides  with a ring plane and the $Z$ axis is directed along the ring axis.
There is another natural Cartesian coordinate system that exists in our theory. In this system, $X0Y$ plane coincides  with the planar surface under consideration.
To avoid ambiguity, we will use the indices $(1, 2, 3)$ to denote coordinates in this system. As usual, the correspondence $(1, 2, 3) \leftrightarrow (X, Y, Z)$ takes place for the coordinate axes.

Equation (\ref{LIE_pert}) describes the dynamics of a vortex filament as a purely geometric object.
But because we are constructing a model to describe the vortex motion, we must take into account the motion of the surrounding fluid in some way.
To do it,   we fulfill two steps.

Firstly, we  assume that 
  the conventional formula for the canonical momentum ${\bf p}$ is fulfilled: 
	\label{VM_1}
	     \begin{equation}
        \label{p_can}
        {\bf p} ~=~ {\bf p}_v ~+~
				\frac{\varrho_0}{2 }\!\int\!\bf{r}\times\bf{w}(\bf{r})\,dV \,,
        \end{equation} 
	where the vorticity 	${\bf{w}}(\bf{r})$ is defined for the vortex filament by stanard way \cite{Nemir}:								
\[{\bf{w}}(\bf{r}) =  \Gamma\!\!
                  \int\limits_{0}^{2\pi}\!\delta(\bf{r} - \bf{r}(\xi))\partial_\xi{\bf{r}}(\xi)d\xi	\,.\]					
   The summand ${\bf p}_v$ takes into account the  momentum which arises when the surrounding fluid has a non-zero velocity $\bf{v}$ without vortex motion.
	The second summand  in the  Eq. (\ref{p_can})   is the standard  vortex canonical momentum ${\bf p}$ for zero boundary conditions $(\bf{v}=0)$ at infinity \cite{Batche}. 
	If we substitute the solution (\ref{our_sol}) into Eq.(\ref{p_can}) we deduce   simple  equation
\begin{equation}
        \label{p_Gamma}
	{{\bf p}} ~-~ {\bf p}_v  ~=~    \pi\varrho_0 {R}^2 \Gamma\, {\bf b}\,. 
	\end{equation}
	The vector ${\bf b}$ here means the binormal unit vector of the considered circular filament;  its coordinates $(b_x, b_y, b_z)$ are $(0,0,1)$.
	It is clear that  this vector coincides with the direction of the vortex circle propagation.

	Secondly, 	we include the circulation $\Gamma$ as a dynamic variable in our theory. 
	Thus, the minimal set of variables needed to describe our dynamic system is as follows:
\[{\cal A} ~=~ \{ \Gamma,\, {\bf q}, \, R, \phi(\tau), \,{\bf b}\}\,.\]

Let us focus our attention on the summand 	${\bf p}_v$ in the Eq.(\ref{p_can}).
		In general, we assume that the following equality is fulfilled:
	\begin{equation}
        \label{p_v_con}
				v_i    ~=~   \sum_{j=1,2}[{\sf M_{eff}}^{-1}]_{ij}({\bf p}_v)_j	\,, \qquad
				i = 1,2\,,
							\end{equation} 	
		where the values  $v_j$ are components of the velocity $\bf{v}$ of the surrounding fluid without vortex motion.  We assume also that the vector $\mathbf{v}$ is parallel to the considered plane surface, so that the fluid flow along the surface.
		The values $[{\sf M_{eff}}^{-1}]_{ij}$ are components of the tensor of the ''inverse effective mass'' for the vortex ring. 
		To calculate its components, we need to determine the energy  $E = E({\bf p},{\bf p}_v)$ of the quantum vortex ring in question.
		Indeed,
	\begin{equation}
	\label{M_def}
	[{\sf M_{eff}}^{-1}]_{ij} ~=~  \frac{\partial^{\,2} E({\bf p},{\bf p}_v)}{\partial p_i\partial p_j}\,,
	\qquad 			i,j = 1,2\,.
	\end{equation}
	More interesting cases for us are those when
	  the vector  ${\bf p}  =  {\bf p}_{sp}$ is an extreme or a stationary point of the function $E({\bf p},{\bf p}_v)$.
 We will return to this issue later.

Therefore, our subsequent goal is to quantize our circular-shaped  vortex filament and   determine its energy.
To quantize our system in accordance with our approach, we need to redefine the set of dynamical variables $\mathcal{A}$.

 Firstly,  we fulfill replacement 
$\{ \Gamma, {\bf b}\} \to  {\bf p}$
with help of the Eq.(\ref{p_Gamma}) where the vector ${\bf p}_v$ is considered as an external parameter.
Secondly,  we replace  the variables $ R$ and  $\phi(\tau)$ with new variables 
$\chi$  and $\varpi$, as follows:
\[ \chi  = \frac{\Delta R}{R_f}\cos(\phi_{\,0} +\omega\tau)\,, \quad  \varpi  = \frac{\Delta R}{R_f}\sin(\phi_{\,0} +\omega\tau) \,, 
\quad \text{where}\quad  \Delta R = \sqrt{R^2 - R_f^2}\,.\]
Clearly, the behavior of the variables $\varpi$ and  $\chi$  is similar to that of a harmonic oscillator.
Finally, we  postulate the set
\begin{equation}
\label{new_set1}
{\mathcal A}^{\,\prime} ~=~ \bigl\{\, {\bf p}\,,  {\bf{q}}\,;\ \varpi\,, \chi\,
  \,\bigr\}
 \end{equation}
as the set of fundamental independent variables for the considered dynamical system -- circle-shaped vortex ring which evolves in accordance with Eq.(\ref{LIE_pert}).
As a consequence of equality (\ref{p_Gamma}), the following formula holds:
\begin{equation}
\label{main_con}
({\bf p} ~-~ {\bf p}_v)^{\,2}   ~=~  \pi^2\varrho_0^2 {R_f}^{\,4}\Bigl(1 + \varpi^{\,2} + \chi^{\,2} \Bigr)^{2}\, \Gamma^{\,2}\,.
\end{equation}
This formula shows that the circulation $\Gamma$ is the function of new variables that parametrize the set (\ref{new_set1}).

\subsection*{3. Hamiltonian structure and quantization}

The structure of the set $\mathcal{A}'$ as well as the assumption about vortex energy
(see  Eq.(\ref{E_general})) naturally lead to the following Hamiltonian description of the model.
\begin{itemize}
  \item Phase space ${\mathcal H} =  {\mathcal H}_{pq}  \times  {\mathcal H}_b   $. The space $ {\mathcal H}_{pq}$ is the phase space for a $3D$  free structureless  particle which     is  parametrized by the variables 
   ${\bf{q}}$ and  ${{\bf p}}$.
	Note that the coordinate $q_3$ cannot equal zero due to the presence of an impermeable planar surface\,\footnote{In this study, we consider the boundary layers both in the domain of $q_3 > 0$ and in the domain of $q_3 < 0$ similarly.}.
	The space    $ {\mathcal H}_b$  is a phase space for one-dimensional harmonic oscillator with  (conditional) frequency $\omega/t_0$.
		 \item Poisson structure:
  \begin{equation}
  \{p_i\,,q_j\}  =  \delta_{ij}\, \quad (i,j = 1,2,3)\,; 
  \qquad \quad
  \{  \varpi,\, \chi\}  =  {1}/{\cal E}_0 t_0 \nonumber 
  \end{equation}
	the brackets that are not written are vanished;
  	\item Hamiltonian 
\begin{equation}
        \label{hamilt_1}
				H ~=~ \frac{{\bf p}^2}{2 \mu_0} ~+~  \frac{{\cal E}_0\,\omega}{2}\Bigl(\varpi^2  + \chi^2  \Bigr)\,.
	\end{equation}
	\end{itemize}

 This Hamiltonian structure  naturally leads to a quantum description of our vortex filament using the  following Hilbert space:
\begin{equation}
	\label{space_quant}
	\bf{H}_1  ~=~  \bf{H}_{pq} \otimes   \bf{H}_b  \,.
	\end{equation}
			{~ Symbol   $\bf{H}_{pq}$  denotes  the Hilbert space  of  quantum states for a free, structureless, non-relativistic particle.  In general, we assume that
			\[\bf{H}_{pq}  ~=~  L_2(\mathbb{R}_2) \otimes L_2(\mathbb{R}_1) 			~\approx~  L_2(\mathbb{R}_3)\,.\]
			Here we limit ourselves to operators on Hilbert space $\bf{H}_{pq}$  of the form				 
			\[ A ~=~  A_2 \otimes I_1 ~+~ I_2 \otimes A_1\,,\]
			where operators $A_2$, $A_1$ and unit operators $I_2$, $I_1$ act in the spaces $L_2(\mathbb{R}_2)$, $L_2(\mathbb{R}_1)$
			correspondingly.  To simplify formulas, we will use notations $A_2 + A_1$ instead $A_2 \otimes I_1 + I_2 \otimes A_1$.
			For our subsequent purposes, we will only consider the $A$-operators with domains:
						\[\mathbb{D}(A) ~\subseteq~ L_2(\mathbb{R}_2) \times L_2(\mathbb{R}_1) ~\subset~
						L_2(\mathbb{R}_2) \otimes L_2(\mathbb{R}_1)\,.\]
											The symbol $\times$ means the algebraic tensor product of the corresponding spaces: any functions $\psi(q_1,q_2,q_3) \in L_2(\mathbb{R}_2) \times L_2(\mathbb{R}_1)$ are  finite sequences of the functions  $\psi_a(q_1,q_2)\psi_b(q_3)$, where $\psi_a(q_1,q_2) \in L_2(\mathbb{R}_2)$ and
	$\psi_b(q_3)\nolinebreak \in \nolinebreak		L_2(\mathbb{R}_1)$\nolinebreak
	\footnote{In the theory of Hilbert spaces, the algebraic tensor product is sometimes called the tensor product, while the tensor product is called the Hilbert tensor product.}.					
			In our case, the domain of any operator $A_1$ must  takes into account the presence of an impenetrable $(X0Y)$ surface.	 Therefore,  			
 any operator $A_1$ which corresponds to a physical observable is defined in some domain
				$ \mathbb{D}_1(A_1) \subseteq
				~\mathbb{D}_1  \subset L_2(\mathbb{R}_1)$, where domain $\mathbb{D}_1$ is defined as follows:
			\begin{equation}
	\label{domain_D} 
		\Psi(q_3) \in 	\mathbb{D}_1: \quad  \Psi(q_3) \in L_2(\mathbb{R}_1)\,, \quad | \quad \Psi(0) =0 \,.
		\end{equation}	
					As it is known, such limitations frequently cause problems with self-adjointness  of   operators.  We will return to the determination of  domain 		$\mathbb{D}({H})$ of the Hamiltinian $H$  later.}		
				
				The symbol $\bf{H}_b$ represents the Hilbert space of the quantized harmonic oscillator, characterized by the classical variables $\chi$ and $\varpi$. 
								This space is formed by the vectors
		\[|\,n\rangle   ~=~  \frac{1}{\sqrt{n!}} (\hat{b}^+)^n    |\,0_b\rangle   \qquad 
		[\,\hat{b}, \hat{b}^+] ~=~ \hat{I}_b\,, \quad \hat{b}|\,0_b\rangle ~=~ 0\,, 
		\label{n_osc} \]
			where vector  $|\,0_b\rangle \in \bf{H}_b$ is vacuum vector and symbols $\hat{b}^+$ and $\hat{b}$ mean the  operators which create and annihilate the oscillator levels. 	
As regards to the classical variables $\varpi$ and $\chi$, these become operators as follows:\label{sigma}
\begin{equation}
	\label{quant_rules}
	\chi ~+~ i\varpi ~\longrightarrow~ \sqrt{2}\,\sigma_{ph}\,\hat{b}\,, \qquad 
	\text{where} \qquad 
\sigma_{ph} ~=~ \sqrt{\frac{\hbar}{{\mathcal E}_0 t_0}}   \,.
\end{equation}
The variables $\bf{q}$ and ${\bf p}$ are quantized using standard rules of quantum mechanics for free, non-relativistic particles in coordinate representation:
$\bf{q} \to \hat{\bf{q}} = \bf{q}$, ${\bf p} \to \hat{{\bf p}}$.  The features of the operator $\hat{{\bf p}}$, which arise due to the presence of an impenetrable surface will be discussed in the subsequent analysis of quantum classical equations.

The first significant finding is that the radius of a vortex ring is quantized. By examining the definitions of the variables $\chi$ and $\varpi$,  we can derive the formula for the quantized value  $R^{\,2}$: 
\[ R^{\,2} ~\longrightarrow~ \hat{R}^{\,2} ~=~ R_f^{\,2}
\left[{\hat I}_b + \sigma_{ph}^2\left({\hat b}^+ {\hat b} + \frac{1}{2}\right)\right]\,.\]   
It is easy to find the eigenvalues  $R^{\,2}_n$  of this operator, and consequently, the  values  ${R}_n$ of the vortex radius:
\begin{equation}
\label{spectrum_R}
 {R}_n ~=~ R_f\sqrt{1 + \sigma_{ph}^2\left(n + {1}/{2}\right)} \,,
\qquad  n ~=~ 0,1,\dots, \,.
\end{equation}
We will consider the spectrum (\ref{spectrum_R})  for a restricted number of $n$ values. As the number \( n \) grows, the system exhibits quasi-classical
 behavior\,\footnote{The author has explored this issue thoroughly in their  work \cite{Tal25_J}.}. 

Further, the quantization scheme results in two spectral problems.
\begin{itemize}
\item  The first spectral problem arises after quantizing the equation (\ref{main_con}). This problem determines the circulation values $\Gamma$:
\begin{equation}
\label{eq_sp_Gamma0}
  \hspace{-5mm}
\left[ \widehat{({{\bf p} - {\bf p}_v})}^{2}  ~-~ \pi^2 \varrho_0^2 \Gamma^2 R_f^4  \Bigl({\hat I}  ~+~  \sigma_{ph}^2\,
(\hat{b}^+ \hat{b} + 1/2)  \Bigr)^2\right]|\Psi\rangle = 0\,.
	\end{equation}
	\item Second problem is determining the energy value $E^\#$, such that the corresponding Hamiltonian
   $\widehat{H}^\#$ provides dynamics in accordance with conditional time $t^\# =  \tau t_0$:
\begin{equation}
	\label{H_quant}
	\widehat{H}^\# |\psi\rangle ~=~ E^\#|\psi\rangle\,, \qquad
		\widehat{H}^\#  ~=~ -~  \frac{\hat{\bf p}^{\,2}}{2\mu_0} ~+~ \frac{\hbar\, \omega}{t_0}
		\left(b^+ b + \frac{1}{2}\right)\,.
		\end{equation}
\end{itemize}

Let us now discuss  operators $\hat{\bf p}^{\,2}$  and $\widehat{({{\bf p} - {\bf p}_v})}^{\,2}$. In the case where the impenetrable $X0Y$ surface is absent, the quantization is trivial:  ${\bf p}^{\,2} \to - \hbar^{\,2} \Delta_3$, where notation $\Delta_n$  ($n =2,3$)	means the Laplace operator on space 	$L_2(\mathbb{R}_n)$ {~ with domain $W^2_2(\mathbb{R}_n)$, where the space $W^2_2(\mathbb{R}_n)$ is  corresponding Sobolev space.}	The quantization of the variable ${({{\bf p} - {\bf p}_v})}^{2}$ is also trivial.

But the operator $\Delta_3$, with a domain {~that   was restricted }
 by Eq. (\ref{domain_D}), is not a self-adjoint operator! Indeed,
\[{\bf p}^{\,2} ~\to~  \hat{\bf p}^{\,2} ~=~ -~\hbar^2 \Delta_2  ~+~ \hat{p}_3^{\,2}\,,\]
where the operator  {~ $\Delta_2$ is self-adjoint operator with domain $\mathbb{D}(\Delta_2) = W^2_2(\mathbb{R}_2)$. Let us discuss the $\hat{p}_3^{\,2}$ operator.  The operator  $\hat{p}_3^{\,2} = -\hbar^2 \partial^{\,2}/\partial {q_3}^2 $  is self-adjoint if it is defined on a domain $W^2_2(\mathbb{R}_1)$. But in accordance with our assumption  (\ref{domain_D}), we must consider the restriction 
$$ \hat{p}_3^{\,2}  ~\longrightarrow~  \check{p}_3^{\,2} ~=~  \hat{p}_3^{\,2}\,\Big\vert_{\mathbb{D}_1 \cap W^2_2(\mathbb{R}_1) }\,,$$
that means the domain restriction
 $$ W^2_2(\mathbb{R}_1) ~\longrightarrow~  \mathbb{D}(\check{p}_3^{\,2}) =  W^2_2(\mathbb{R}_1) \cap \mathbb{D}_1\,.$$
The operator $\check{p}_3^{\,2}$ is symmetric operator in the space $L_2(\mathbb{R}_1)$,
so that the inclusion $\mathbb{D}(\check{p}_3^{\,2}) \subset \mathbb{D}\bigl((\check{p}_3^{\,2})^+\bigr)$ is fulfilled.
Therefore, we must consider the self-adjoint extension $\check{p}_3^{\,2}(\lambda)$ (see, for example, \cite{ReedSim2}) of this operator. 
To avoid technical details that are not relevant to the main topic of this study, we will use the results from work \cite{FilShaf}.
In accordance with results \cite{FilShaf}, the all self-adjoint extensions  $\check{p}_3^{\,2}(\lambda)$  of the operator $\check{p}_3^{\,2}$
are marked by the real number $\lambda$.  The corresponding  domains $\mathbb{D}\bigl(\check{p}_3^{\,2}(\lambda)\bigr)$
includes the functions $\psi (x) \in \mathbb{D}\bigl((\check{p}_3^{\,2})^+\bigr)$ that meet specific conditions: 
\begin{eqnarray}
\psi(-0) ~-~ \psi(+0)  &=& 0\,,\nonumber\\
\psi^\prime(-0) ~-~ \psi^\prime(+0)  &=& - \frac{2\lambda}{\hbar^2}\psi(0)\,.\nonumber
\end{eqnarray} 
Formally speaking, these functions can be considered as the solution of the differential equation
\[ \left[-~ \frac{\hbar^2}{2\mu_0} \frac{\partial^{\,2}}{\partial {q_3}^2} ~+~ \lambda \delta({q_3})\right]\psi(q_3) ~=~   E_\lambda \psi(q_3)\,, \qquad   E_\lambda < 0\,,\]
where $\lambda = - \hbar\sqrt{-2 E_\lambda/\mu_0}$. For example, the function 
\begin{equation}
\label{wave_f}
\Psi_\lambda({q_3}) ~=~ \left\{ \begin{matrix}
~~\exp[ - \sqrt{-2\mu_0 E_\lambda}\, {q_3}/\hbar]\,, &  ~~{q_3} > 0\,,\\[3mm]
~~\exp[ ~ \sqrt{-2\mu_0 E_\lambda}\, {q_3}/\hbar]\,, &  ~~{q_3} < 0\,,
\end{matrix}\right.
\end{equation}
solves this equation quite correctly: the product $\delta({q_3})\psi(q_3)$ is well-defined here.
This fact gives some basis for interpreting of the  extension procedure described above as a background for the Schr\" odinger equation with a $\delta$ potential.  In $3D$ case, this fact was investigated many years ago \cite{BerFad}.

Of course, the differential expression $[-~ ({\hbar^2}/{2\mu_0}) ({\partial^{\,2}}/{\partial {q_3}^2}) + \lambda \delta({q_3})]$ with
the distribution-valued coefficient function 
does not define the operator in the space $L_2(\mathbb{R}_1)$. As a consequence, treating this differential equation as the Schr\" odinger equation is problematic.  To do this, we need to construct a theory in which the multiplication of distributions is a well-defined prosedure.
For example, corresponding theory was constructed in the work \cite{Shir}. We do not go into detail, but we note that the solution (\ref{wave_f}) has also been deduced from the  theory,  suggested in the paper \cite{Shir}. 

To simplify notations, we will use the summand  $2\mu_0\lambda\,\delta({q_3})$ formally, meaning that the procedure
of the self-adjoint extension of the operator  $\check{p}_3^{\,2}$  was fulfilled.}


Thus,  Eq.(\ref{eq_sp_Gamma0}) is transformed as follows:
\begin{eqnarray}
\label{eq_sp_Gamma}
  \Biggl[ -~ \hbar^2 \biggl(\frac{\partial}{\partial {q_1}} ~-~ ({\bf\sf k}_v)_1 \biggr)^2  
&-& \hbar^2 \biggl(\frac{\partial}{\partial {q_2}} ~-~ ({\bf\sf k}_v)_2 \biggr)^2 
 -~ \hbar^2\frac{\partial^{\,2}}{\partial {q_3}^2}  ~+\\[2mm]
 ~+~ 2\mu_0\lambda\,\delta({q_3})
&-& \pi^2 \varrho_0^{\,2}\, \Gamma^2 R_f^{\,4}\,  \Bigl({\hat I}  ~+~  \sigma_{ph}^2\,
(\hat{b}^+ \hat{b} + 1/2)  \Bigr)^2\Biggr]\,|\Psi\rangle = 0\,, \nonumber
	\end{eqnarray}
where the values $({\bf\sf k}_v)_i$, where $i=1,2$,  are the components of the wave vector  ${\bf p}_v/\hbar$.
  Spectral problem
(\ref{eq_sp_Gamma}) has eigenvectors
\begin{equation}
\label{eig_1}
| \Psi \rangle ~=~ |\Psi_\lambda(p_1,p_2)\rangle |\,n \rangle\,,
\end{equation}
where the vectors $|\Psi_\lambda(p_1,p_2)\rangle  \in    \bf{H}_{pq}^\prime $ in coordinate representation take form:
\[ \langle {\bf q}\,|\,\Psi_\lambda(p_1,p_2)\rangle ~\equiv~ \Psi(q_1,q_2,q_3) ~=~
\Psi_\lambda({q_3})\exp[ -i (p_1 {q_1} + p_2 {q_2})/\hbar^2]\,.\] 
Symbol $\bf{H}_{pq}^\prime$ means corresponding rigged (''equipped'') Hilbert space (see, for example, \cite{BerShu}).

\subsection*{4. Why is it a boundary layer?}

Thus, we achieved one of our goals in this study:  the vectors (\ref{eig_1})  visually demonstrate that the vortex in question is located near the $Z=0$ surface. 
The eigenvalues $\Gamma^{\,2}$  of the spectral problem (\ref{eq_sp_Gamma}) determine  the permitted
circulation values in our theory. 
To write $\Gamma$ values in a convenient form, we will introduce some additional notations:
\[   
{\sf M}_f ~=~ \pi \varrho_0 R_f^{\,3}\,, \quad p_{\hbar} ~=~ \frac{\hbar}{R_f}\,, \label{Scale_p}
 \quad \boldsymbol{\kappa} ~=~
\frac{{\bf p}}{p_{\hbar}}\,, \quad  \boldsymbol{\kappa}_v ~=~
\frac{{\bf p}_v}{p_{\hbar}}  ~=~ R_f \bf{\sf k}_v \,.\]

 The value $p_{\hbar}$ defines the momentum scale in our theory. As regards the mass parameter ${\sf M}_f$, its physical meaning will be clarified later.
 Further, we will prefer dimensionless vectors, $\boldsymbol\kappa$, $\boldsymbol \kappa_v$, \dots, anywhere if it does not cause ambiguity.

	As a consequence of the above studies, the values of circulation are written in terms of "mass"  ${\sf M}_f$ and dimensionless vectors $\boldsymbol{\kappa}$, $\boldsymbol{\kappa}_v$ as follows:

\begin{equation}
\label{Circ_1}
\Gamma_\lambda({\bf p},{\bf p}_v;n) ~=~ \pm~ \frac{\hbar}{{\sf M}_f}
\gamma_{\lambda}({\bf p},{\bf p}_v;n)\,,
	\end{equation}
	where dimensionless function $\gamma_{\lambda}({\bf p},{\bf p}_v;n)$ is written as follows:
	\[\gamma_{\lambda}({\bf p},{\bf p}_v;n) ~\equiv~  \gamma_{\lambda} (\boldsymbol{\kappa},\boldsymbol{\kappa}_v; n) ~=~ 
	\frac{\sqrt{(\kappa_1 - (\boldsymbol{\kappa}_v)_1)^2 ~+~
  (\kappa_2 - (\boldsymbol{\kappa}_v)_2)^2   ~-~  2\beta_\lambda /\sigma_{ph}^2}}{ [1 ~+~ \sigma_{ph}^2(n + 1/2)]}\,.\]
	The  notation $\beta_\lambda = R_f |E_\lambda|/v_0 \hbar$ was introduced here.
	\label{beta_lambda}

This formula requires further discussion. We won't discuss  here the notable discrepancies between the found set of circulation values and the standard set $\Gamma \propto k$, where the number $k$ is natural number. The author's previous works delve deeply into these disparities, examining specific cases and their corresponding arguments and motivations.
The new feature of the system under discussion is the absence of a $\Gamma$-value for some parameter values. Indeed, we can see that the function  under the square root may be negative, but the circulation must always be a real number.

To better understand the physical implications of this observation, we will begin with a qualitative analysis.
Let us suppose that the matrix 

${\sf M_{eff}}^{-1} \nolinebreak =  \nolinebreak ([{\sf M_{eff}}^{-1}]_{ij})$ exists and  the inverse matrix ${\sf M_{eff}}$ also   exists, such that each element $[{\sf M_{eff}}]_{ij}$  is finite.  Thus, we consider a ''easy vortices'' here.
In this case, the condition for the square root of Eq.(\ref {Circ_1} ) to take non-zero real values is as follows:
\begin{equation}
\label{real_cond1}
\sum_{i,j = 1}^2 \left( p_i - [{\sf M_{eff}}]_{i,j}{v}_j\right)^2 ~>~
 2\mu_0|E_\lambda|\,.
\end{equation}
It is clear that this condition will break down in one of two ways.
First one, when the  values of $\bf{v}$ and ${\bf p}$ are sufficiently small.  Second one, when the values of $\bf{v}$ and ${\bf p}$ can be  arbitrarily  large but the  following condition holds:
\begin{equation}
\label{criterion1}
\Big\vert\,p_i - \sum_{j = 1,2}[{\sf M_{eff}}]_{i,j}{v}_j\,\Big\vert ~\le~ \sqrt{2\mu_0|E_\lambda|}
\,,\qquad \,.
\end{equation}
 These  facts means that there is no any vortex motion in this case. 
Thus, the prediction of our model agrees quite well with the paper \cite {StPaBa} where the creation of vortices in a boundary  layer is conditioned by irregularities in the surface. As it seems, the ''almost resting'' fluid should not lead to vortex formation. 
As regards the condition (\ref{criterion1}), it predicts the disappearance of vortex motion in a certain range of fluid velocities.  
We're still examining individual circular  vortices. Of course, the turbulent media in the boundary layer consists of many vortices of various shapes and sizes. 
We'll return to this issue in subsequent sections of this study.

The above qualitative analysis is quite conditional. Indeed,
when the matrix ${\sf M_{eff}}$ ceases to exist for certain parameter values, the criterion for the vortex appearance  becomes more complex. In this case, we must consider the equations:
\begin{eqnarray}
\label{sys_1}
|\,{\bf p} ~-~{\bf p}_v|^{\,2} &>& 2\mu_0|E_\lambda|\,,\\[2mm]
\label{sys_2}
v_i  &=& \sum_{j=1}^2\frac{\partial^{\,2} E({\bf p},{\bf p}_v)}{\partial p_i\partial p_j}\,     ({\bf p}_v)_j\,,  \qquad  i = 1,2\,.
\end{eqnarray}
To deduce the condition of the vortices appearance,  we need to find the  values $({\bf p}_v)_1$  and $({\bf p}_v)_2$ from the system (\ref{sys_2}).  After that,  the result should be 
substituted  into Eq.(\ref{sys_1}).


Thus, to continue our study, we need to return to the definition of the vortex energy  $E({\bf p},{\bf p}_v)$  in our theory.
The spectral problem (\ref{H_quant}) has the eigenvalues
\begin{equation}
	\label{E_eigen}
	E^{\#}_{n}({\bf p}) ~=~ \frac{1 }{2\mu_0 }  \Bigl(p_1^{\,2} + p_2^{\,2}\, \Bigr) ~+~ E_\lambda   ~+~
	\frac{\hbar\,\omega}{t_0}\, \left(n + \frac{1}{2}\right)\, \qquad n = 0,1,\dots\,.
	\end{equation}
	
	As an important point, we must remember that until now we have described LIA dynamics using the concept of ''conditional time'' - a special evolution parameter, $ t^* = t \Gamma / 4 \pi $.  		
		After that, we introduced   the dimensionless evolution parameter $\tau$  in accordance with Eq.(\ref{tau_s}) for convenience. 
This fact means that the values (\ref{E_eigen} ) represent some "conditional energy", since this  energy value is related to the  dynamics with respect to "conditional" time $t^\# = t_0\tau$ .
Therefore, our first goal is to return to the actual (physical) time $t$ as well as to return to
actual  energy $E_{n}(p_1,p_2)$, which is connected to the  time $t$.
To do it, let us note that the $t^\#$ dependence of any vector $|\Psi\rangle \in H_1$ is described as follows:
		\begin{eqnarray}
	\label{t_ev1}
	~&~&~|\Psi\rangle ~\longrightarrow~ |\Psi(t^\#)\rangle ~=~ 
		\exp\left(\frac{i\widehat{H}^\# t^\#}{\hbar} \right)|\Psi\rangle ~= \nonumber\\[3mm]
				~&=& \sum_{n} \int\!\! dp_1 dp_2 \, C_{n}(p_1,p_2)\exp\left(\frac{iE^{\#}_{n}({\bf p}) t^\#}{\hbar} \right)
		|\Psi_\lambda(p_1,p_2)\rangle |\,n \rangle\,, \nonumber
		\end{eqnarray}
	where  $ C_{n}(p_1,p_2) = (\langle\Psi_\lambda(p_1,p_2)|\langle n|)  |\Psi\rangle$.	In the next step,
					we  restore the real-time $t$ value in our formulas by using the equation
				\begin{equation}
	\label{t_real}
		 t^\# ~\to~  t ~=~ \frac{4\pi R^2}{ t_0 |\Gamma|}\,t^\#\,.
		\end{equation}	
							
			Our key assumption here is that the probabilities  
							$|\langle \Phi  |\Psi(t^\#)\rangle |^2$ of transition  to any quantum  state $|\Phi\rangle$  	are observable values that do not depend on the time parametrization. 
					To ensure this, we need to  require 	the following equality  for the real energy $E_{n}(p_1,p_2)$: 
				\[  E^{\#}_{n}(p_1,p_2)\, t^\#  ~=~  E_{n}(p_1,p_2)\, t\,.\]

Therefore, 				the ''real - time'' evolution of any vector $|\Psi\rangle \in {\bf H}$   is written as
		\begin{equation}
	\label{t_ev2}
		|\Psi\rangle ~\longrightarrow~ |\Psi(t)\rangle ~=~  
		\sum_{n} \int\!\! dp_1 dp_2\, C_{n}(p_1,p_2)\exp\left(\frac{iE_{n}(p_1,p_2)\, t}{\hbar} \right)
		|\psi_{{p};\,\lambda, n}\rangle\,,
		\end{equation}
					where function $E_{n}(p_1,p_2) \equiv  E_{n}({\bf p}; {\bf p}_v) $ means the actual vortex energy.				
This function takes the following form:
\begin{equation}
	\label{E_true}
	 E_{n}({\bf p}; {\bf p}_v)  ~=~ \frac{t_0 |\,\Gamma_\lambda({\bf p},{\bf p}_v;n)\,|}{4\pi R^2_n}E^{\#}_{n}({\bf p}) ~\equiv~
		E_{\hbar}\,{\cal E}_{n}({\boldsymbol\kappa}, {\boldsymbol\kappa}_v)\,,
	\end{equation}
	where the constant
	\[E_{\hbar} ~=~ \frac{\hbar^2}{ {\sf M}_f R_f^2}\]
	defines the energy scale  in our model. The dimensionless function ${\cal E}_{n}({\boldsymbol\kappa}, {\boldsymbol\kappa}_v)$
	is written as follows:
	\begin{eqnarray}
	\label{E_dim_less}
{\cal E}_{n}({\boldsymbol\kappa}, {\boldsymbol\kappa}_v) &=& \frac{\sqrt{(\kappa_1 - 
({\boldsymbol\kappa}_v)_1)^2  + (\kappa_2 - ({\boldsymbol\kappa}_v)_2)^2 -
 2 \beta_\lambda/\sigma_{ph}^2}}{ 4\pi\bigl[ 1 +  \sigma_{ph}^2(n + 1/2)\bigr]^2}~\times\nonumber\\[2mm]
&\times& \left[ \omega\bigl(n + 1/2\bigr) + \frac{\sigma_{ph}^2}{2} \bigl(\kappa_1^2 +  \kappa_2^2   \bigr)  - \beta_\lambda\right]\,.
\end{eqnarray}

Let us discuss the breakdown of the condition (\ref{sys_1}) from the function ${E}_{n}(\boldsymbol{\kappa}, \boldsymbol{\kappa}_v)$  point of view. Indeed, the breakdown of this condition for some values $(\kappa_1, \kappa_2)$
 means that this function  takes the imaginary  values. We will denote corresponding domain in the $(\kappa_1, \kappa_2)$ plane as $D_{00}$.
 Let the condition (\ref{sys_1})  be broken. 
By writing the corresponding appropriate square root value in the form 
\[ {E}_{n}({\boldsymbol\kappa}, {\boldsymbol\kappa}_v) ~=~ {i\,\hbar}/{T_{n}({\boldsymbol\kappa},{\boldsymbol\kappa}_v)}\,, \qquad T_{n}({\boldsymbol\kappa},{\boldsymbol\kappa}_v) ~>~ 0\,,\]
we  can transform some parts of a sub-integral expression in Equation (\ref{t_ev2}) into a form:
\[\exp\bigl({iE_{n}(p_1,p_2)\, t}/{\hbar} \bigr)
		|\psi_{{p};\, \lambda, n}\rangle  ~\longrightarrow~ 
\exp\bigl( -~ {t}/{T_{n}({\boldsymbol\kappa},{\boldsymbol\kappa}_v)} \bigr)
		|\psi_{{p};\, \lambda, n}\rangle\,.\]
Thus, the vortex modes corresponding to the given values of $(\kappa_1, \kappa_2)$ are unstable. Their contribution decreases exponentially over time for any given amplitude. 
 In this context, the constant $T_n({\boldsymbol\kappa}, {\boldsymbol\kappa}_v)$ can be linked to a "vortex lifetime." In a classical theory, this parameter has been studied in the work \cite{Tenn}, from a different point of view.
 
The above outcome is consistent with the previous statement that vortices do not exist if condition (\ref{sys_1}) is broken.
More specific results that justify the applicability of our model to a boundary layer description,  require a detailed study of the function  ${E}_{n}(\boldsymbol{\kappa}, \boldsymbol{\kappa}_v)$.

\subsection*{5. The energy structure of vortices and the conditions for their formation}

The first obvious conclusion is that the tensor of the inverse effective mass (\ref{M_def})   contains off-diagonal elements.
To continue our study, we will investigate the structure of the $3D$ surface $\mathcal{E}_n = \mathcal{E}_{n}(\kappa_{1}, \kappa_{2}; \boldsymbol{\kappa}_{v} )$ in the Cartesian coordinate system  $({\cal E}, \kappa_1, \kappa_2)$.  
One of the important issues that we intend to explore is the dependency of the set of 
the  extremal and  stationary points of the function  $\mathcal{E}_{n}(\kappa_{1}, \kappa_{2}; \boldsymbol{\kappa}_{v} )$ on the  parameter  
$\boldsymbol{\kappa}_{v}$.
To avoid the imaging $3D$ graphics which will not sufficiently visually demonstrate the structure of the ${\cal E}_{n}(\boldsymbol{\kappa}; \boldsymbol{\kappa}_{v})$ function,  we consider the structure of the levels of this function.
It turns out that the picture of these levels  essentially  depends  on the vector 
 ${\bf p}_{v} = p_{\hbar}  \boldsymbol{\kappa}_v$ as well as other  parameters of our model.
We must consider two cases.

$\star$  The case when $\omega (n + 1/2) ~>~ \beta_\lambda\,$.
In this case, the square brackets in Eq.(\ref{E_dim_less}) 
do not vanish for any values of vector  $\boldsymbol{\kappa}$ and remain  positive.

\begin{center}
		 {\includegraphics[width=6.0in]{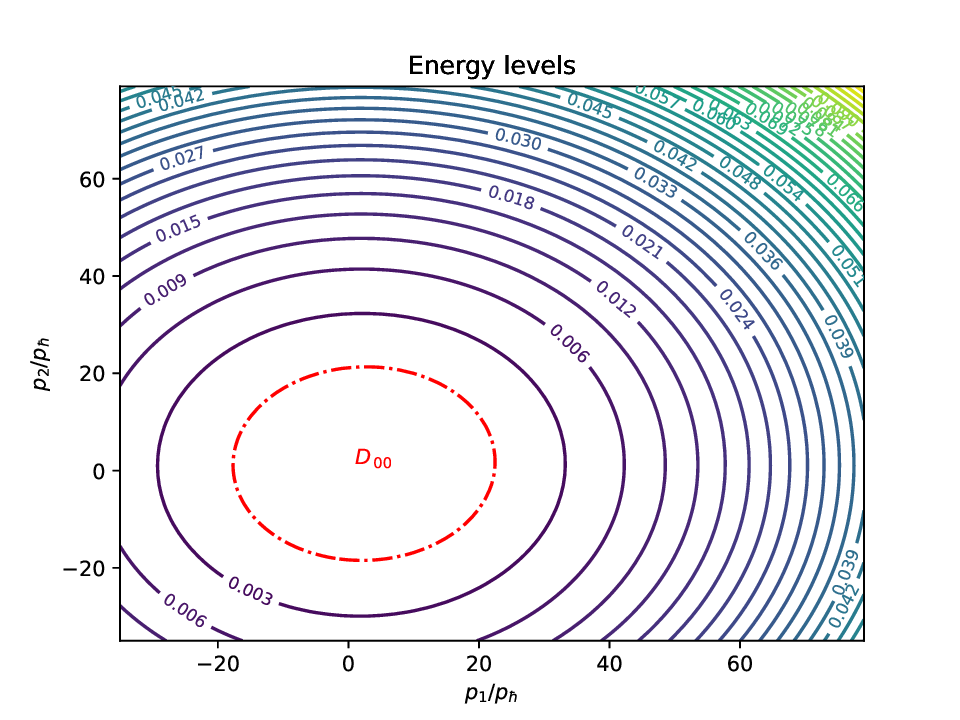}}
{\captionof{figure}{Energy levels for $(\boldsymbol{\kappa}_v)_1= 5$, $(\boldsymbol{\kappa}_v)_2= 3$.
\label{p_v_small}}}
\end{center}

So, when the value of $|{\boldsymbol\kappa}_{v}|$ is sufficiently small, we get a simple level structure that is demonstrated in Fig.\ref {p_v_small}.
The domain $D_{00}$   where the condition (\ref{sys_1}) is broken, is bounded in Fig.\ref {p_v_small} by the  red  dashed curve.  At  each point on this curve,  the function ${\cal E}_{n}(\boldsymbol{\kappa}, \boldsymbol{\kappa}_v)$ reaches a  local minimum which is equal to zero.

\begin{center}
		 {\includegraphics[width=6.0in]{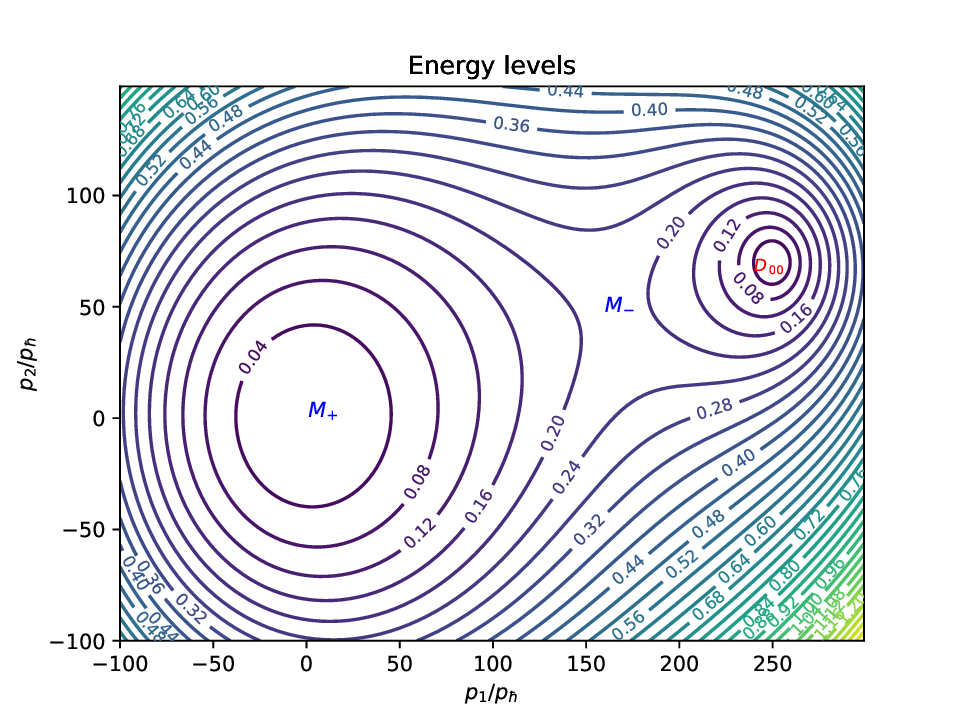}}
{\captionof{figure}{Energy levels for $(\boldsymbol{\kappa}_v)_1= 250$, $(\boldsymbol{\kappa}_v)_2= 70$.
\label{p_v_large}}}
\end{center}

What happens when ${\cal E}_{n}(\boldsymbol{\kappa}; \boldsymbol{\kappa}_v) > 0$ in the case of
small value of  $|{\boldsymbol\kappa}_{v}|$?
 To find out,   let us view a simple formula
\[\frac{\partial {\cal E} }{\partial t } ~=~ \frac{\partial {\cal E} }{\partial p_1 } \dot{p}_1 ~+~   \frac{\partial {\cal E} }{\partial p_2 }\dot{p}_2\,, \qquad  \dot{p}_i \equiv 
\frac{d p_i }{d t }\,.\]

Still,  we consider the integrable Hamiltonian system (\ref{LIE_pert}), so that the equalities
$\dot{p}_i = 0$ hold.  But really, in many vortex systems, we must consider the various interactions between vortices.  In this scenario, $\dot{p}_i$ generally does not equal zero.
Therefore, the any stable (or quasi-stable) vortex state corresponds only to a stationary point, where the equalities ${\partial {\cal E} }/{\partial p_i } =0$,
$i = 1,2$ take place.
The case under study the surface $\mathcal{E}_n = \mathcal{E}_{n}(\kappa_{1}, \kappa_{2}; \boldsymbol{\kappa}_v)$ has not any stationary points.  Consequently, vortices do not form here and the flow of fluid in the boundary layer remains laminar.

When the value of $|{\boldsymbol\kappa}_{v}|$ increases, the level picture becomes more complex. 
Increasing  the value of $|{\boldsymbol\kappa}_{v}|$, the domain $D_{00}$ also moves so that it is located in some neighborhood  of the point $\bigl((\boldsymbol{\kappa}_v)_1, (\boldsymbol{\kappa}_v)_2\bigr)$. 
When the number $|{\boldsymbol\kappa}_{v}|$ reaches and exceeds
  some critical value  \label{Pv_cr}
${\boldsymbol\kappa}_{cr} = {\bf p}_{cr}/p_{\hbar}$, the surface ${\cal E} =  {\cal E}_{n}(\boldsymbol{\kappa}, \boldsymbol{\kappa}_v)$ gets two additional local extreme points -- $M_+$ and $M_-$ where the equalities  ${\partial {\cal E} }/{\partial p_i } =0$, $i = 1,2$, hold.
 The level structure  for this case  is demonstrated in Fig.\ref {p_v_large}.
The extreme points -- $M_+$ and $M_-$  have positive and negative curvature correspondingly.
Critical point ${\boldsymbol\kappa}_{v} =  {\boldsymbol\kappa}_{cr}$ is characterized by the following condition being fulfilled: \label{P_0}
\begin{equation}
\label{crit_point}
\exists\, {\bf p}_0 = p_{\hbar}  \boldsymbol{\kappa}_0 \qquad \Big\vert \qquad
\frac{\partial^{\,2} {\cal E}_{n}(\boldsymbol{\kappa}, \boldsymbol{\kappa}_{cr})}{\partial{\kappa}_i\,\partial{\kappa}_j}\,\bigg\vert_{\boldsymbol{\kappa} = \boldsymbol{\kappa}_0}  ~=~ 0
\qquad  \forall\, i,j = 1,2\,.  
\end{equation}
It is clear that a planar point occurs at a moment when the picture in Fig.\ref {p_v_small}  transforms  into the  picture in Fig.\ref {p_v_large}, by means of  increasing of the value of ${\boldsymbol\kappa}_{v}$.
Thus, we can only observe the vortices appearing near the plane $Z=0$ when
${\bf p}_v =p_{\hbar}\boldsymbol{\kappa}_v  >  {\bf p}_0$.
Formally, taking into account Eqs.(\ref{p_v_con}) and (\ref{crit_point}), we can see that
 this domain for the parameter ${\bf p}_v$ corresponds to the domain 
$\bf{v}>0$ for the fluid velocity $\bf{v}$. It is clear that the vanishing of the matrix ${\sf M^{-1}_{eff}}$ corresponds to the vortex with ''infinitely large mass''.
We will explore this surprising phenomenon later in this section.

$\star$  The case when $\omega (n + 1/2) ~\le~ \beta_\lambda\,$.
In this case, the square brackets in  Eq.(\ref{E_dim_less}) 
take negative  values for $\boldsymbol{\kappa}  \in  D_{neg}$
and they are vanish for  values  of  $\boldsymbol{\kappa}  \in \partial D_{neg}$: 
\[D_{neg}: \qquad \{ \boldsymbol{\kappa}\in \mathbb{R}_2 \quad |\quad  {\sigma_{ph}^2} \bigl(\kappa_1^2 +  \kappa_2^2   \bigr)   ~<~  2\beta_\lambda ~-~ \omega\bigl(2n + 1\bigr) \,\}\,.\]

\begin{figure}[h]
{
\begin{minipage}[h]{0.49\linewidth}
\center{\includegraphics[width=3.0in]{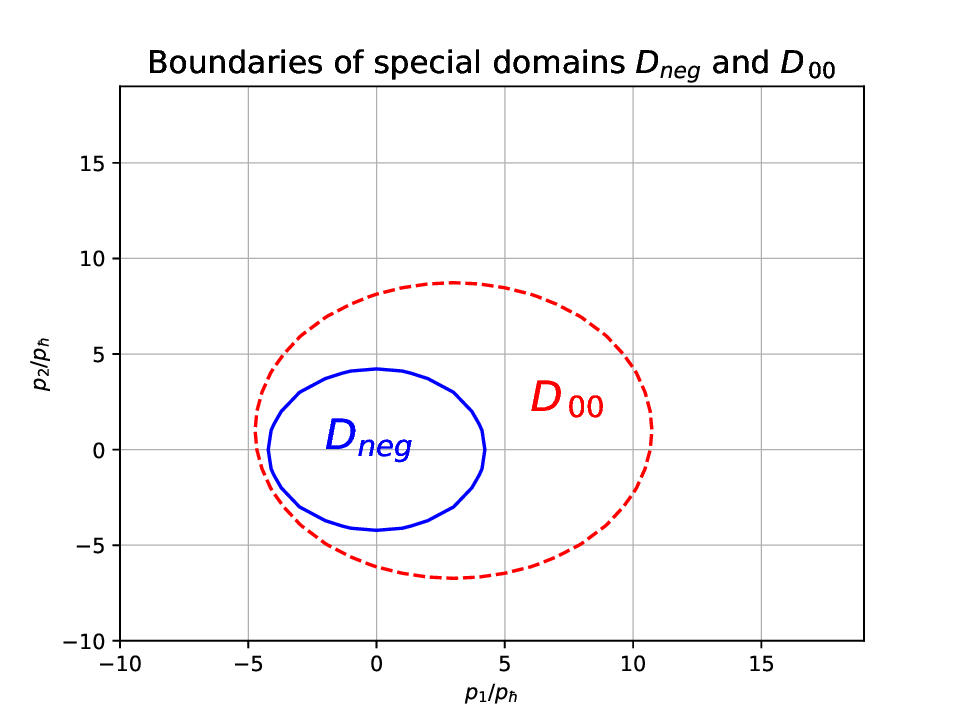} }
{\captionof{figure}{Domain's boundaries \\  for $(\boldsymbol{\kappa}_v)_1= 3$, $(\boldsymbol{\kappa}_v)_2= 1$.
\label{DD_small}}}
\end{minipage}}
\hfill
{
\begin{minipage}[h]{0.49\linewidth}
\center{\includegraphics[width=3.0in]{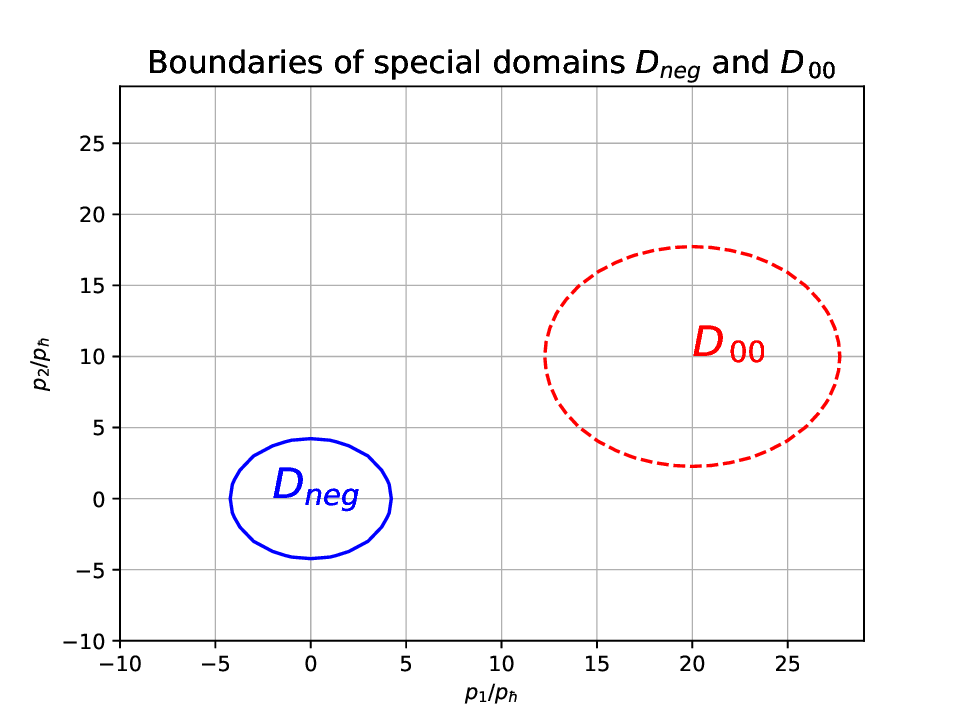} }
{\captionof{figure}{Domain's boundaries \\ for $(\boldsymbol{\kappa}_v)_1= 20$, $(\boldsymbol{\kappa}_v)_2= 10$.
\label{DD_large}}}
\end{minipage}}
\end{figure}

The relative location of the domains $D_{neg}$ and $D_{00}$ depends essentially on the vector 
$\boldsymbol{\kappa}_v$.
If the value $|\boldsymbol{\kappa}_v|$ sufficiently small, the inclusion
$D_{neg} \subset D_{00}$ takes place. This case demonstrated on the Fig.\ref{DD_small}.
Thus, we have the same case as was visualized in Fig.\ref{p_v_small}, because the energy becomes an imaginary value at all inner points in the domain $D_{00}$.

As $|{\boldsymbol\kappa}_{v}|$ grows, the center of the domain $D_{00}$ shifts along with the vector ${\boldsymbol\kappa}_{v}$, while the domain $D_{neg}$ stays near the origin $(0,0)$. 
This case demonstrated on the Fig.\ref{DD_large}.

Thus, we have a local energy minimum that takes negative values in this case.

Our analysis demonstrates that the point ${\bf p} = {\bf p}_0$, where the ''inverse effective mass'' tensor vanishes, exists.
Nevertheless, the existence of the vortices with an infinite masses  seems unphysical.
Let us analyze additional restrictions imposed on our system by quantum mechanics.

\begin{itemize}
 \item Let us consider the case when the local extreme points $M_\pm$ exist (case where  $|{\bf p}_{v}| > |{\bf p}_{cr}|$)   but the
   equality  $|{\bf p}_v - {\bf p}_{cr}| < \epsilon$   holds for a small value of $\epsilon$. Let the extreme points $M_\pm$   have  coordinates  ${\bf p}_{\pm}$.
	\label{M_pl_min}   Then,  the energy ${E}_{n}({\bf p}, {\bf p}_{v})$ in  small neighbourhoods of the points  ${\bf p}_{\pm}$  can be written 
in the form
\begin{equation}
\label{1_energy}
{E}_{n}({\bf p}, {\bf p}_{v}) ~-~ {E}_{n}({\bf p}_\pm, {\bf p}_{v})  ~\simeq~ \sum_{i,j =1}^2
\frac{\partial^{\,2} {E}_{n}({\bf p}_{\pm}, {\bf p}_{v})}{\partial{p}_i\,\partial
{p}_j}\,({\bf p}  -{\bf p}_{\pm})_i ({\bf p}  -{\bf p}_{\pm})_j\,.
\end{equation}
Because our assumptions, the second derivatives in the Eq.(\ref{1_energy}) take small values.

\item When we study the domains of physical values of $E$ and ${\bf p}$  near certain fixed points, we must take into account the Heisenberg uncertainty principle.  
In actual quantum systems, the uncertainties $\Delta E$  and $\Delta{\bf p}$  in measuring energy and momentum are never zero.
So, the assumption $|\Delta{\bf r}|_{min} \simeq R_f$ for the  minimal 
uncertainty $|\Delta{\bf r}|$  of coordinate ${\bf r}$ seems quite true.
Therefore, the corresponding uncertainty  $|\,\Delta{\bf p}\,|_1$ that minimizes the value
$|\Delta{\bf r}|\,|\Delta{\bf p}\,|$ will be roughly the order of $p_{\hbar}$.

As regards of the the uncertainty $\Delta E$, we assume that $|\,\Delta E\,| \simeq E_{\hbar}$. 
Indeed, in accordance with Eq.(\ref{E_true}), 
 we use a ''ruler'' with divisions of $E_{\hbar}$ for the energy determination.
Correspondingly, the precision of any measurement will not be higher than one division. 
Therefore, we can assume that in a small neighbourhood of the points $\bf p_{\pm}$, the estimating equalities hold:
\begin{eqnarray}
|\,{E}_{n}({\bf p}, {\bf p}_{v}) ~-~ {E}_{n}({\bf p}_\pm, {\bf p}_{v})\,|   &\simeq& |\,\Delta E\,| ~\simeq~ E_{\hbar}\,, \nonumber\\
|\,{\bf p}  ~-~ {\bf p}_{\pm}|  &\simeq& |\,\Delta{\bf p}\,|_{1}
~\simeq~  p_{\hbar}\nonumber
\end{eqnarray}

\item   Applying the Schwarz inequality to Equation (\ref{1_energy}), the estimates from above yield an inequality
\[  |\,{E}_{n}({\bf p}, {\bf p}_{v}) ~-~ {E}_{n}({\bf p}_\pm, {\bf p}_{v})\,| ~\le~ \Vert {\sf M_{eff}^{-1}}\Vert\,
 |\,{\bf p}  ~-~ {\bf p}_{\pm}\,|^2\,,\]
where symbol $\Vert {\sf M_{eff}^{-1}}\Vert$ means the norm of inverse effective matrix
$ {\sf M_{eff}^{-1}}$. Thus, quantum mechanics, together with the specific features of the dynamical system being studied, leads to the following restriction on the matrix 
${\sf M^{-1}_{eff}}$:

\[\Vert {\sf M_{eff}^{-1}}\Vert ~\ge~  \frac{1}{ \sf M_{f}}  \,.\]
 This inequality clarifies the physical meaning of the mass constant $\sf M_{f}$ which was introduced at the beginning of the section 4. This constant defines the upper limit of the effective mass of the quantum  vortices  under study:
\[  \Vert {\sf M_{eff}}\Vert ~ \le~ { \sf M_{f}}\, \text{Cond}({\sf M_{eff}})\,,\]
\end{itemize}
where the multiplier $\text{Cond}({\sf M_{eff}})$ is the condition number for the matrix ${\sf M_{eff}}$.
Looking back at the Eq.(\ref{p_v_con}), we are once again convinced that vortices do not form in a  ''slowly flowing fluid''.

\subsection*{6. Concluding remarks: from a single quantum  vortex to  the boundary layer formation}

Sinse we considered a single quantum vortex ring near the impenetrable plane.
The proposed model demonstrates a higher probability of the location of such vortices near the surface. 
Discussing turbulent motion in the boundary layer, we need to consider the many-vortex system. The simplest system here consists of a number of non-interacting or incredibly  weakly interacting circular vortices with meso- and micro- sizes.
The state space of such a quantum system is as follows:
\[\bf{H}_K ~=~  \underbrace{\,\bf{H}_1 \otimes\dots\otimes\bf{H}_1\,}_{K}\,,\]
where Hilbert spaces $\bf{H}_1$ are defined as in Eq.(\ref{space_quant}) for each vortex individually.
{~ We assume the vortices follow Bose statistics, so we consider a symmetric tensor product.}

The full Hamiltonian takes the form
\[   {\widehat H}_v ~=~ \sum_{k=1}^K  \underbrace{I \otimes I \otimes\dots\otimes}_{k-1} {\widehat H}(\lambda_k, \omega_k) \underbrace{\otimes I \otimes \dots\otimes I}_{K - k} \,,\]
where,  {~ by virtue of spectral theorem,   }
\[{\widehat H}(\lambda_k, \omega_k) ~=~ \sum_{n} \int\!\! dp_1 dp_2 E_{n}({\bf p}; {\bf p}_v) |\,\psi_{{p};\,\lambda_k, n}\rangle\langle \psi_{{p};\,\lambda_k, n}\,|\,.\]
 {~  In accordance with our previous constructions, operators    ${\widehat H}(\lambda_k, \omega_k)$ are self-adjoint operators.
This means that full Hamiltonian ${\widehat H}_v$ is self-adjoint operator.}
Parameters both $\lambda_k < 0$ and  ''dimensionless frequencies'' $\omega_k$ {~ may vary from vortex to vortex.}
   The vortices in a turbulent flow are short-lived structures \cite{Tenn}.   Because they are     located near the different points of the plane $Z=0$, the set of the numbers $\lambda \in [ \lambda_{min}  \lambda_{max} ]$   models the  irregularities of the surface.  The permitted interval $[ \lambda_{min}  \lambda_{max} ]$ as well as the distribution law of this parameter should  depends on the  type of surface.  This issue is a separate study item. 
The vortex configurations that are more complex than rings can be considered using the method suggested by the author in the paper \cite{Tal26_1}.

The above studies demonstrate the existence of stable minima in the vortex energy  (point $M_+$ on Fig.\ref{p_v_large}), which can take positive and negative values depending on the values of the parameters $\lambda_k$ and $\omega_k$. 
This means that a transition to a turbulent regime in the boundary layer does not imply a jump-like energy change in a quantum system.
As regards the states which allow us to interpret our system as a quasi-particle with negative effective mass (point $M_-$ on Fig.\ref{p_v_large}), it could be valuable for simulating superfluidity in quantum turbulent media. The author hopes to return to this topic in future works.

\subsection*{7. Appendix}

In order to facilitate the understanding of the graphs and their corresponding conclusions, we present in the table some of the notation used.

\vspace{5mm}

{\footnotesize 

\begin{tabular}{|c||l|c|} \hline 
                         &                                   &   \sf The  page     \\
 ~~\sf\normalsize{Notation} &~~~~\sf\normalsize What does this value mean?  & \sf where the value \\
                &             &\sf  was defined \\ \hline\hline
\normalsize${p}_{\hbar}$  &  Scale factor which has &  page \pageref{Scale_p}     \\
                   &   the dimension of momentum     &       \\ \hline
 \normalsize ${\bf p} ~\equiv~ {\boldsymbol\kappa} p_{\hbar}$  & Full vortex momentum  & page \pageref{VM_1} \\
 	               &          &                     \\ \hline
 \normalsize ${\bf p}_v~\equiv~ {\boldsymbol\kappa}_v p_{\hbar}$ &  Additional    vortex momentum caused by  & page \pageref{VM_1} \\ 
		                  &    the movement of the surrounding fluid  &      \\ \hline
 \normalsize ${\bf p}_{cr}~\equiv~ {\boldsymbol\kappa}_{cr} p_{\hbar} $  & Critical value of
 ${\bf p}_v$ when the planar & page \pageref{Pv_cr}  \\
                 &  point on graph ${\mathcal E}_{n}({\boldsymbol\kappa}, {\boldsymbol\kappa}_v)$ appears  &  \\ \hline
\normalsize ${\bf p}_0 ~\equiv~ {\boldsymbol\kappa}_0 p_{\hbar}$  & Coordinates of the planar point, where  &  page \pageref{P_0} \\
                   &  components of $\partial^2{\mathcal E}_{n}/\partial p_i \partial p_j$ vanish &    \\ \hline
\normalsize ${\bf p}_\pm ~\equiv~ {\boldsymbol\kappa}_\pm p_{\hbar}$  & Coordinates of the local stationary & page \pageref{M_pl_min} \\ 
           &  points $M_\pm$ which appear when  $|{\bf p}_v| >  |{\bf p}_{cr}|$    &   \\ \hline
 \end{tabular}}

\vspace{5mm}

Below, we provide the constants and quantum numbers employed in creating the graphs.

\vspace{5mm}

\begin{tabular}{|c|c|c|c|} \hline 
                      &   &     &   \sf The  page   \\
 ~~\sf{Constant} & Figures 1,\,2 &  Figures 3,\,4 & \sf where the value  \\
            &    &   & \sf  was defined  \\ \hline\hline
						& & &    \\
\Large $\sigma^2_{ph}$ &\Large $10^{\,-6}$ &\Large  $10^{\,-6}$ &page \pageref{sigma}\\
                   &       &   &    \\ \hline
		& &  &   \\ 
	\Large	$\beta_{\lambda}$  & \Large $2 \cdot 10^{\,-5}$& \Large $3 \cdot 10^{\,-5}$&  page \pageref{beta_lambda}\\
		& &  &   \\    \hline	
		& &  &   \\						
  \Large $\omega$  & \Large $10^{\,-4}$ & \Large $14\cdot 10^{\,-6}$& page \pageref{omega} \\
 	               &          &   &             \\ \hline
 \sf Quantum  &  &  &  \\
\sf number         &   1    & 1& page \pageref{n_osc} \\    
\Large $n$  & & &  \\ \hline
   \end{tabular}

\vspace{5mm}

The   constants $v_0$, $\varrho_0$,   $R_f$, $\mu_0$     do not directly influence both  the ''dimensionless circulation''  $\gamma_{\lambda}({\bf p},{\bf p}_v;n)$  and the ''dimensionless energy'' 
 ${\cal E}_{n}(\boldsymbol{\kappa}, \boldsymbol{\kappa}_v)$. Their effects are mediated solely through the values of the constants  $\sigma_{ph}$ and $\beta_{\lambda}$. 
As noted earlier, the constants $\alpha$, $\omega$ and $E_{\lambda}$, 
 can vary from one vortex to another.

\end{document}